\documentclass[useAMS,usenatbib]{mn2e}
\usepackage{url}
\usepackage{aas_macros}
\usepackage{amssymb}
\usepackage{amsmath}
\usepackage{graphicx}
\usepackage{bm}
\usepackage{multicol}
\usepackage{subfig}
\usepackage[usenames]{color}
\usepackage{float}
\usepackage{psfrag}
\usepackage{fixltx2e}
\newcommand{\alert}{\textcolor{black}}
\newcommand{\ha}{{\rm H}$\alpha$~}
\newcommand{\ly}{{\rm Ly}$\alpha$~}

\title[Scattering of emission lines in galaxy cluster cores]
{Scattering of emission lines in galaxy cluster cores:
  measuring electron temperature}

\author[Khedekar et al.]{
\parbox[t]{16cm}{
S. Khedekar$^{1}$\thanks{satej@mpa-garching.mpg.de}, E. Churazov$^{1,2}$\thanks{churazov@mpa-garching.mpg.de}, S. Sazonov$^{2,3,1}$, R. Sunyaev$^{1,2}$,\\ E. Emsellem$^{4,5}$
}\\
\\
$^{1}$MPI f\"ur Astrophysik, Karl-Schwarzschild str. 1, Garching, 85741, Germany\\
$^{2}$Space Research Institute, Profsoyuznaya str. 84/32, Moscow, 117997, Russia\\
$^{3}$Moscow Institute of Physics and Technology, Institutsky per. 9, 141700 Dolgoprudny, Russia\\
$^{4}$European Southern Observatory, Karl-Schwarzschild-Str. 2, 85748 Garching, Germany\\
$^{5}$Universit´e Lyon 1, Observatoire de Lyon, Centre de Recherche Astrophysique de Lyon
and \\ \ Ecole Normale Sup´erieure de Lyon, 9 avenue Charles Andr´e, F-69230 Saint-Genis Laval, France\\
}

\begin{document}

\date{Draft: \today}

\pagerange{\pageref{firstpage}--\pageref{lastpage}} \pubyear{2013}

\maketitle

\label{firstpage}

\begin{abstract}

The central galaxies of some clusters can be strong emitters in the \ly and \ha lines. This emission may arise either from the
 cool/warm gas located in the cool core of the cluster or from the bright AGN within the central galaxy. The luminosities of
 such lines can be as high as $10^{42} - 10^{44}$ erg/s. This emission originating from the core of the cluster will get Thomson
 scattered by hot electrons of the intra-cluster medium (ICM) with an optical depth $\sim$ 0.01 giving rise to very broad ($\Delta \lambda
/ \lambda \sim$ 15\%) features in the scattered spectrum. We discuss the possibility of measuring the electron
 density and temperature using information on the flux and width of the highly broadened line features.
\end{abstract}


\section{Introduction}

Centres of most relaxed galaxy clusters host giant visibly
prominent elliptical galaxies referred to as the brightest cluster
galaxy (BCG). A few of these BCGs show significant star-formation
rates, $\sim 10-100 \ {\rm M}_{\odot}/{\rm yr}$, \cite[but also as
  large as $\sim 740 \ {\rm M}_{\odot}/{\rm yr}$,
  see][]{2012Natur.488..349M}.  As tracers of cool/warm gas, bright
emission lines have been observed at the centres of clusters across a
wide range of wavelengths. For example Ly$\alpha$
\citep{1992ApJ...391..608H, 1995MNRAS.273..625J, 2005ApJ...632..122B}
and $[$OII$]$ \citep{1989AJ.....98.2018M} in UV, \ha+$[$NII$]$
in optical \citep{1981ApJ...250L..59H, 1983ApJ...272...29C,
  1999MNRAS.306..857C}; rotational lines from molecular H$_2$
 \citep{1994ASSL..190..169E, 1997MNRAS.284L...1J, 2000ApJ...545..670D} and
CO \citep[][]{2001MNRAS.328..762E, 2003A&A...412..657S};
$[$NeII$]$ and polycyclic aromatic hydrocarbons (PAH) complexes \citep{2011ApJ...732...40D}.
 Recent Herschel observations \citep[][]{2011MNRAS.418.2386M, 2012MNRAS.426.2957M, 2012A&A...542A..32C}
 have also detected atomic cooling lines of $[$CII$]$, $[$OI$]$, $[$NII$]$,
 providing us with important clues about the thermal state of gas in cluster
 cores.

It is also well known that the centres of BCG host an Active Galactic Nucleus (AGN) which is powered by the accretion
 of mass onto the central super-massive ($10^6 - 10^9 {\rm M}_{\odot}$) black hole. These AGN's can
 be extremely luminous across a broad range of frequencies from the Radio to $\gamma$-rays. These
 bright emission lines in the AGN spectra arise as a result of photoionisation of the cold/warm gas
 by the continuum radiation coming from the centre. Some of the many prominent lines are
 \ly, CIV, CIII], MgII, NeV, [OI], [OII], [OIII], [NeII], \ha, H$\beta$, H$\gamma$, [NII], etc.,
see \cite{AGN1, AGN2, AGN3} for reviews.

The aim of this paper is to highlight the diagnostic potential of the
bright emission lines originating from the centres of clusters to
 measure the ICM properties. These lines should get highly
broadened after being Thomson scattered by the hot electrons ($T_e
\sim$ few keV) of the ICM (with optical depth, $\tau \sim
0.01$). The width of these broadened lines would be a probe of
the line-of-sight (LOS) weighted gas temperature, while the ratio of
the flux-density in the scattered and direct continua, would give in
optical depth or the LOS-weighted gas density. Thus the ability to
 measure this scattered signal will open up an alternate and interesting
 probe of the ICM.

The observations of scattered light from clusters has been discussed
previously in the literature in various contexts. The possibility of
observing beamed AGN radiation scattered by the surrounding gas was discussed
by \cite{1982SvAL....8..175S, 1987SvAL...13..233G}.
The observed alignment between radio and optical emission (in the 3C/3CR radio galaxies), first reported by
 \citet{1987Natur.329..604C}, was attributed to the scattering of the beamed emission of the AGN \citep{1989MNRAS.238P..41F} by hot electrons
 from the intra-cluster/group gas. The evidence for scattering was further supported by polarimetric
 observations \citep{1989Natur.341..307D}. However, subsequent works \citep[see e.g.,][]{1996ApJ...465..157D, 2005AJ....129.1212Z}
 ruled out scattering by hot ($T_e \gtrsim 10^4$ K) electrons, as constrained by the observed lines widths,
 instead preferring the scattering by relatively cool gas and dust. \cite{2002A&A...393..793S}
considered resonant scattering of X-ray emission from a central AGN by the ambient intracluster gas, proposing it as
 a method to indicate the parameters of the surrounding hot gas.

Also, \cite{2004ApJ...602..659H} proposed that
joint observations of scattered radio emission along with the SZ
effect could be used to probe the temperature profiles of
clusters. The SZ effect probes the quantity $y=\int n_e T_e dl$ while the scattered
flux from point sources in radio would probe optical depth $\tau =
\sigma_T \int n_e dl$; assuming a $\beta$-model for the gas density,
they show that temperature measurements could be possible up to a
precision of 1 keV. The scattering and polarisation of light at
optical wavelengths was discussed by \cite{1993ApJ...418...60M},
however they did not consider the possibility of using the spectral
information to measure temperatures.

The outline of this paper is as follows. In section
\ref{sec:kernel} we illustrate the scattering of a single bright line
and show the dependence of the broadened profile on the electron
temperature. In section \ref{sec:perseus} we consider the scattering
of \ha lines in Perseus cluster and compute the surface brightness of
the scattered signal as a function of the radius. In section \ref{sec:AGN}
 we repeat a similar computation, but now for \ly and \ha lines emitted by the
  AGN in the bright radio galaxy 3C 295 found within a galaxy cluster. In section
\ref{sec:feas} we discuss the feasibility of measuring this faint and
diffuse scattered signal and outline the requirements for such
observations. We also comment on the possibility of exploiting the
 polarisation information of the scattered signal in the outer envelopes
 of cD galaxies / intracluster light to improve
the chances of detecting the scattered signal. In section
\ref{sec:discuss} we mention the optimal properties of clusters needed
 for the robust measurement of temperature and optical depths in
clusters. We conclude in section \ref{sec:conclude}.

\section{Broadened spectral profile of a narrow bright line}
\label{sec:kernel}

The Thomson scattering of photons by non-relativistic (\alert{$k_B
T_e/m_e c^2 \ll 1$}), electrons having a Maxwellian distribution at temperature
$T_e$ is described by a photon redistribution kernel,
 given by \cite[see][]{1970JQSRT..10.1277B, 1980SvAL....6..213S},
\begin{align}
& K(\nu, \bm{\Omega} \rightarrow \nu', \bm{\Omega}') =& \nonumber \\
& \frac{3}{16\pi} \sqrt{\frac{m_e c^2}{2\pi k_B T_e}} \left( \frac{\nu'}{\nu} \right) \frac{1+\mu^2}{g}
 \exp{ \left\lbrace -\frac{m_e c^2 (\nu-\nu')^2}{2 k_B T_e g^2} \right\rbrace}&,
\label{eqn:kernel}
\end{align}
where $\mu=\bm{\Omega} \cdot \bm{\Omega}'$ is the cosine of the angle of
scattering and $g=\sqrt{\nu^2 +\nu'^2 -2 \nu \nu' \mu}$. Note that
this kernel is applicable for single scattering approximation.
Throughout this paper we shall consider multi-scattering to be negligible,
which is a valid assumption for an a optically thin gas (in clusters,
$ \tau=\sigma_T \int_{0}^{\infty}n_e(r)dr \ll 1 $).
\alert{At IR, optical, UV and soft X-ray frequencies and for temperatures $k_B T_e \leq 10 ~ {\rm keV}$ equation
 \ref{eqn:kernel} is sufficiently accurate, and we shall use this form in the rest of the paper.
\cite{2000ApJ...543...28S} provide a more general expression (see their Eq. 7)
 for the kernel $K$, applicable in the mildly relativistic regime, up to
 $k_B T_e \lesssim 25 ~ {\rm keV}$ and $h\nu \lesssim 50 ~ {\rm keV}$.
 Note that \cite{1980SvAL....6..213S} (see Eq. 5)
 and \cite{2000ApJ...543...28S} (see Eq. 19) also
 give an analytic expression for the above formula in the case of angle-averaged
 scattering.}

\begin{figure}
  \centering
  \subfloat[Impact of the intrinsic line width on the angle-averaged scattered profile. Any intrinsic line width causes
 smearing of the cuspy shape at the tip of the scattered line profile. Here $\delta(\lambda-\lambda_0)$ and $G(\Delta \lambda, \lambda_0)$ denote the
 intrinsic shape of the emission line -- the delta function and the Gaussian profile (with ${\rm s.d.} = \Delta \lambda$) respectively.
 Changes in the flux at the wings are seen only when the intrinsic width of the line is comparable/larger than thermal broadening.]
 {\includegraphics[width=0.5\textwidth]{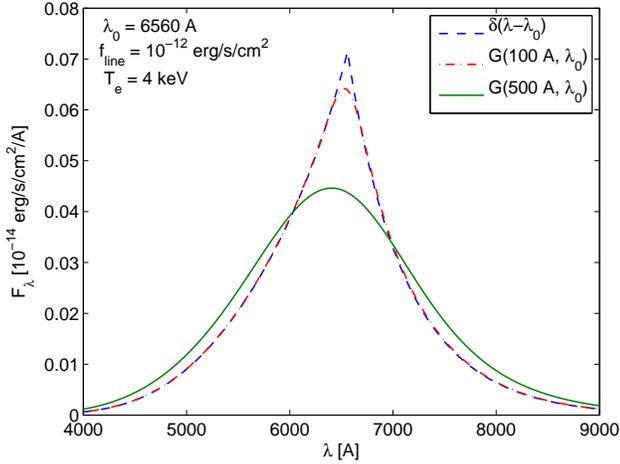}}
  \hspace{5mm}
  \subfloat[Angle-averaged broadened line profile after scattering at various temperatures.
 Note that the multi-temperature-weighted broadening (thin dashed black),
 with the weights $w_1$ and $w_3$ for the temperatures $T_1$ and $T_3$, is not the same as that produced due to a single
 but similarly weighted temperature (thick dash-dotted magenta), $T_2=w_1 T_1 + w_3 T_3$.]
{\includegraphics[width=0.5\textwidth]{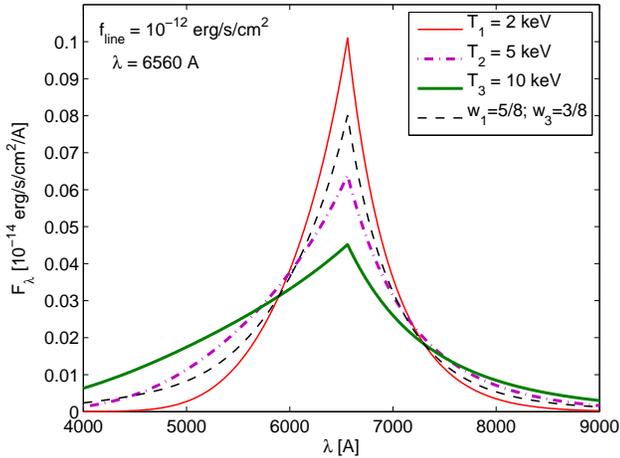}}
  \caption{Broadened line profile due to scattering (after angle-averaging): effects of intrinsic line width and gas temperatures on
 the shape of the scattered line. A flux of $1 \times 10^{-12}$ erg/cm$^2$/s is assumed for the flux in the unscattered line.
 }
  \label{fig:broad}
\end{figure}

Fig. \ref{fig:broad} (a) shows the broadened and angle-averaged spectral profile of a single
line at 6560 A using the kernel in equation \ref{eqn:kernel} and assuming a
temperature of $T_e=4$ keV.  The three curves correspond to the
following intrinsic line shapes: a delta function and two Gaussians with
  standard deviation of 100 and 500 A respectively; all having the same central wavelength and
same flux in the unscattered line. If the  line is intrinsically narrow
the angle-averaged scattered line shows a sharp cusp at the wavelength of the line
with broad asymmetric wings on either sides \alert{\citep{1980SvAL....6..213S}}. 
\alert{
In the Thomson limit, when a photon is scattered by an electron with
velocity $\bm v$ the energy of the former changes as
\begin{equation}
\frac{\nu'}{\nu}=\frac{1-\frac{\bm \Omega \cdot \bm v}{c}}{1-\frac{\bm \Omega'\cdot \bm v}{c}}.
\end{equation}
Naively, this suggests that averaging over an isotropic (e.g. Maxwellian)
distribution of electrons
will lead to a symmetric scattered line profile. However, the rate of
scatterings is proportional to
$(1-\frac{\bm{\Omega \cdot \bm v}}{c})$, i.e. photons are preferentially
scattered by approaching rather than
receding electrons. This effect (also generic to the Fermi type-II
acceleration mechanism) causes a net
change of the photon energy $\propto \left (\frac{v}{c} \right )^2 \propto
kT_e/m_ec^2$ and skews the scattered line profile to the right.
}

A line with a finite width will cause a smoothing of the cusp. For most bright lines of interest the
 intrinsic line width is small compared to the broadening caused by the scattering, i.e.,
\begin{equation}
\frac{\Delta \lambda}{\lambda}  \ll \sqrt{\frac{2 k_B T_e} {m_e c^2}} \sim \frac{4 \times 10^4 ~ {\rm km/s}}{c} .
\end{equation}
For example \citet{2012ApJ...746..153M} find the width of emission lines between 600 km/s (in the nuclei)
 to 100 km/s (in the outer filaments). Thus the wings of the broadened line profile are expected to remain unaffected and any
 change in the shape of the line would occur only near the tip depending on the line width $\Delta \lambda$.

In Fig. \ref{fig:broad} (b) we show the effect of gas temperature on the
broadening. A useful estimate of the width of the broadened line (see
equation \ref{fig:broad}), is $\Delta \lambda/ \lambda \approx
0.2\sqrt{k_B T_e/ 10\ {\rm keV}}$.  At a given projected distance from
the cluster centre, the LOS passes through gas at various
temperatures. The scattered line will then be broadened by different amounts,
 weighted by the product of gas and photon density distribution along the
LOS. Fig. \ref{fig:broad} (b) shows the (angle-averaged) multi-temperature broadening with the
LOS weights of 5/8 and 3/8 for temperatures of 2 and 10 keV. One can
also see that this weighted profile is different from a single
temperature profile with temperature of 5 keV which is the weighted
mean of the two temperatures.

\begin{figure}
 \centering
     \includegraphics[width=0.5\textwidth]{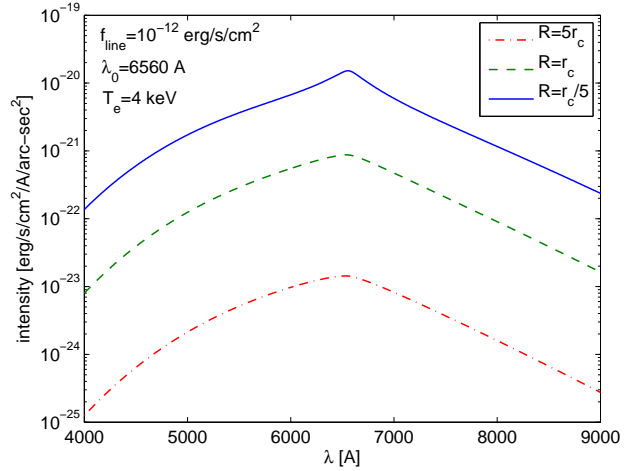}
  \caption{Shape of a single scattered line at various projected distances assuming an isothermal $\beta$-model profile for the Perseus cluster.
 The tip of the profile becomes more cuspy towards the centre due to significant scattering at small angles, while it is flat at
 larger distances where the scattering occurring at angles close to $\pi/2$ becomes dominant.}
  \label{fig:cuspy}
\end{figure}

The brightness/intensity of the scattered light from a central isotropic source in a (spherically symmetric) cluster may be derived as follows.
Let $P(\nu)$ be the luminosity [erg/s/Hz] of the central source producing the flux $f(\nu,r)=P(\nu)/4\pi r^2$ at the scattering point in the cluster.
The surface brightness profile of scattered radiation at the projected distance $R$ is then,
\begin{equation}
I(\nu, R) = \underset{\l} \int \underset{\nu'} \int f(\nu,r) K(\nu' \rightarrow \nu, \mu) n_e(r) \sigma_T d\nu' dl
\label{eqn:los}
\end{equation}
where $r=\sqrt{l^2+R^2}$, $\mu=l/r$, and $l$ is the distance along the LOS.

Fig. \ref{fig:broad} shows a thermally broadened line profile due to the angle-averaged scattering of light (emitted by a central point source)
 from the surrounding gas. One gets a cuspy profile at the tip ($\nu=\nu'$), only when the scattering occurs from regions very close
 to the central source. In reality deviations (from the highly cuspy profile) are expected owing
 to the fact that one would typically be interested in observing the scattered light (say in a circular annulus), excluding the highly bright
 central region emitting directly. Equation \ref{eqn:kernel} shows the dependence of the
scattering kernel on the angle of scattering, $\mu=\cos(\theta)$; note that the function $g(\nu,\nu';\mu) \rightarrow 0$ as $\nu \rightarrow \nu'$ and
 $\mu \rightarrow 1$ (forward-scattering), causing a sharp cusp at the tip. Thus when the scattered light is observed close to the centre in projection
 (when small-angle forward scattering is dominant) the broadened line is cuspy; while for the scattered light observed away from the centre (scattering
 at angles close to $\theta=\pi/2$ dominate) the profile is smooth at the tip. The dependence of the shape of the broadened line
for scattered light observed at various values of projected distances is indicated in Fig. \ref{fig:cuspy}. We emphasise that this
 effect should be considered in deducing the gas temperatures from observations of the scattered lines \cite[also see][]{2000ApJ...543...28S}.

\begin{figure}
  \centering
  \subfloat[Scattered and direct spectrum  of NGC 1275 (in the observer frame). The left axis in {\it (thick) blue}
 indicates the scattered flux in the thermally broadened spectrum (within a radius $R_{\tau/2}=50$ kpc),
 while the right axis in {\it (thin) red} shows the
 intensity of the direct spectrum (from Buttiglione et al. 2009) collected over an aperture of $2''$.
 The colours of the axis correspond to those of the respective spectra.]
 {\includegraphics[width=0.5\textwidth \hspace{-2mm}]{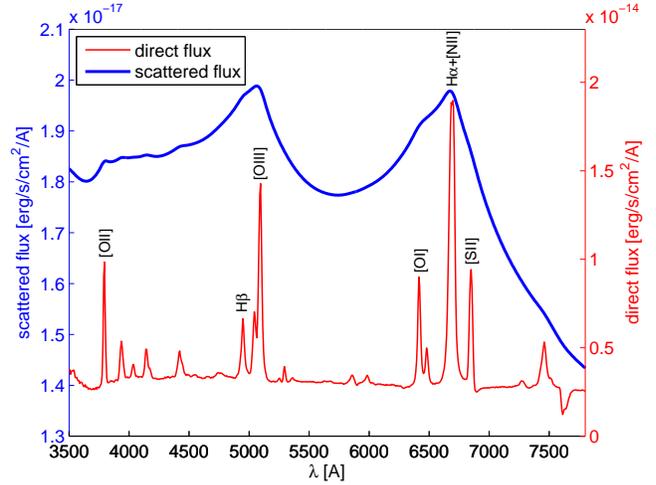}}
 \hspace{3mm}
 \subfloat[Surface brightness of the scattered \ha emission shown in {\it (thick) blue line} as a function of
 the projected distance $R$ from the cluster centre. The {\it (thin) red} line shows the direct (or unscattered)
 brightness profile (measured at 550 nm, but converted to the observed \ha wavelength) of the NGC 1275 stellar continuum
 from Prestwich et al. 1997. For all the lines in this plot, the left axis indicates the AB magnitude while the right axis shows
 the corresponding specific intensity.]
 {\includegraphics[width=0.5\textwidth]{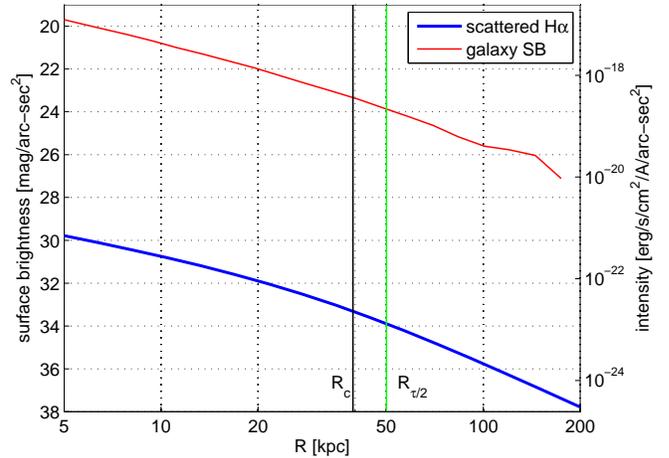}}\\
   \hspace{-5mm}
\subfloat[Expected polarisation of the scattered signal as a function of the projected radius.]
     {\includegraphics[width=0.5\textwidth]{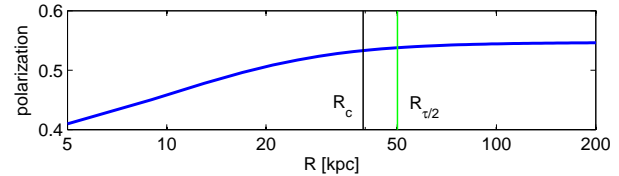}}
  \caption{Estimate of the expected signal of the diffuse scattered
    light around the central galaxy NGC 1275 in Perseus cluster.}
  \label{fig:perseus}
\end{figure}

\section{Broadening of the [OIII] and [NII]+\boldsymbol{${\rm H}\alpha$} emission lines in Perseus cluster}
\label{sec:perseus}

The Perseus cluster (also known as Abell 426) is a massive nearby cluster with a redshift of $z=0.0179$ (comoving distance of 75.3 Mpc).
 It has a cool core with the \ha luminosity\footnote{The value of luminosity quoted here has been adjusted to comply with a currently accepted value
 of the Hubble parameter ($h=0.7$) from $h=0.5$ assumed in \citet{2001MNRAS.328..762E}.} of $2.4 \times 10^{42}$ erg/s
 \citep{2001MNRAS.328..762E}, showing one of the most luminous \ha line among the nearby clusters. The \ha emission
 in Perseus cluster extends beyond its central galaxy, NGC 1275, in the form of bright filaments \citep[occurring up
 to $2'$  from the centre, see][]{2001AJ....122.2281C} which appear to be lifted up by the rising bubbles of hot
 plasma generated by the central AGN \citep{2000A&A...356..788C, 2001ApJ...554..261C}. Prominent \ha filaments are also found
in other cool core clusters \cite[for e.g.,][]{2010ApJ...721.1262M}.

We now estimate the surface brightness and the shape of the
spectrum originating from NGC 1275, after scattering by the hot gas
surrounding it. The temperature and density variations along the LOS
are taken into account (see equation \ref{eqn:los}) while computing
this scattered spectrum; we use the density and temperature profiles from the ACCEPT catalogue
\citep{2009ApJS..182...12C}. The radial Thomson optical depth of the Perseus cluster
(estimated by fitting a $\beta$ model to the ACCEPT data) is $\tau \sim 7.8 \times 10^{-3}$ and
 its core radius, $R_c$, is 39 kpc.

Fig. \ref{fig:perseus} (a) shows the spectrum (in thin red) of NGC 1275 in the
wavelength range 3500-7800 A which was taken by
\cite{2009A&A...495.1033B} with the 3.58 m Italian Telescopio
Nazionale Galileo over an aperture\footnote{\alert{We estimated (from the available HST images in the H$\alpha$ filter)
that the central 2'' region is very bright, containing most of the flux. Including the flux from the outer region
 would increase this value by at most another $\sim$ 25\%. Since an accurately measured flux is available
 only for the central 2'' and the aim of the paper is to make just a feasibility study, we decided to use this
 value to estimate the scattered flux.}} of $2''$. Some of the prominent lines (with corresponding
rest frame vacuum wavelengths in Angstrom) indicated in the spectrum are:
[OII] 3727,3730, H$\beta$ 4863, [OIII] 4960,5008, [OI] 6302,
[NII]+\ha 6550,6585;6565 and [SII] 6718,6733. The
broadened spectrum (in thick blue) is computed by integrating the scattered
spectra within a radius $R_{\tau/2} = 50$ kpc, where the radial
optical depth drops to 50\% of its total value. Note that the
scattered spectrum is much smoother since the contrast between peaks
and continuum is decreased after broadening. The total scattered flux
 is $\tau \sim 10^{-2}$ times smaller than the direct flux.

In Fig. \ref{fig:perseus} (b) the {\it (thick) blue} line shows the surface brightness (AB mag/arc-sec$^2$)
 of the scattered light as a function of the projected distance from the centre of the Perseus
 cluster. This intensity is estimated near the \ha wavelength.
 Although the  plotted line assumes a central point source for the emission, in reality
 the scattered emission would be flatter on scales comparable to the size of the dominant H$\alpha$-emitting region.
 The {\it (thin) red} line shows the surface brightness of the host galaxy continuum at the observed \ha
 wavelength, estimated\footnote{
The conversion of magnitude from 550 nm to the observed \ha wavelength was then done assuming that
 the continuum part of the spectrum is flat and more or less constant in shape. Fig. \ref{fig:perseus} (a)
 shows that this is a reasonable assumption.} from \cite{1997ApJ...477..144P}.
It is clear that the host galaxy continuum from the stellar contribution is always expected to be much brighter than the expected signal
 from scattered light by $\sim$10 magnitudes.

Scattering of light from a central source produces a net polarisation at every point such that the direction
 of the polarisation vector is perpendicular to the radial line joining the given point to the centre \cite[also see][]{1982SvAL....8..175S}.
 This information about the polarisation of the signal would be useful in separating the contribution from the direct stellar continuum of the galaxy
 which would be unpolarised. Of course, the stellar continuum would also be scattered and polarised, but this contribution would be fainter by
 a factor of $\sim \tau$ compared to the direct light.
 Fig. \ref{fig:perseus} (c) shows the expected polarisation degree defined as
 $(I_{\perp}-I_{\parallel})/(I_{\perp}+I_{\parallel})$, where $I_{\perp}$ and $I_{\parallel}$ are the intensities
of the scattered light in planes perpendicular and parallel to the scattering plane.

\section{Scattering of the L\MakeLowercase{y}$\bm \alpha$ and H$\bm \alpha$ lines emitted by an AGN within a galaxy cluster}
\label{sec:AGN}
Obviously the diffuse scattered light originating from the \ha emission in and around the central galaxy of the Perseus cluster is very faint by itself,
 and also in comparison with the surface brightness profile of the central galaxy. The detection of such signal is very challenging, at least in the immediate future.

We now discuss an alternate scenario that would allow us to measure the temperature of the cluster ICM using the \ha and/or Ly$\alpha$ lines arising
 from a bright AGN located in the central galaxy. The central galaxies of clusters often have an AGN that emits across a broad range of frequencies
 from radio to high-energy X-rays, and even Gamma rays, with the presence of several bright emission lines occurring over the entire spectrum.
 
All AGNs are believed to have common features like a supermassive ($10^6 ~{\rm to}~ 10^9 ~{\rm M_{\odot}}$) black hole
 surrounded by an accretion disk and hot corona; the presence of high (inner region) and low (outer region) velocity
 gas, also referred to as the broad and narrow line regions respectively; and in between these two regions usually lies an optically
 thick torus of dust and gas. In addition, a relativistic jet originates from within 100 Schwarzschild radius of
 the black extending upto few 10's of kpc (radio-quiet) to Mpc (radio-loud). According to the unification scheme the
 different observed properties of AGN, like overall spectral energy distribution and the presence or absence of the broad spectral lines,
 arise only due to different orientations of the AGN with respect to the observer. If the torus obscures the emission
 from the broad line region of the AGN it is classified as type II, and otherwise as type I.

If the broad-line emission region is hidden from us, as in a type II AGN, and instead we observe only the scattered light, then we could use
 the information from thermal broadening of the scattered bright emission lines to measure the weighted temperature of the cluster
in which the AGN sits. Powerful type II AGNs (radio loud galaxies), especially of the type FRII, sitting within large galaxy clusters would be
 ideal candidates for making such a measurement. Cygnus A is one such example, being the most luminous nearby AGN, with a
 bolometric\footnote{In the template spectrum that we use (to estimate the scattered signal), the line luminosity increases with the bolometric luminosity.}
 luminosity of $\sim 3 \times 10^{45} \rm{erg/s}$. However it is located close to the galactic plane and hence there would be a strong
 absorption (HI column density along the line of sight is $\sim 3.5 \times 10^{21} {\rm cm^{-2}}$), especially so for the \ly line in the scattered emission.
 Here we consider another powerful AGN sitting in the narrow-line radio galaxy 3C 295 (type II AGN) at a redshift of 0.4605, having a bolometric luminosity
 (quite similar to Cygnus A) of $\sim 2 \times 10^{45} \rm{erg/s}$. This cluster is well suited for our needs as it has a relatively large Thomson optical depth of $1.4 \times 10^{-2}$.

\begin{figure}
  \centering
  \caption{Scattered and direct spectrum (but hidden from the observer),
 near the \ly line, of the AGN in the central galaxy 3C 295 in the observer frame,
 where the optical depth drops to half its total value. The left axis in {\it blue}
 indicates the scattered flux of the thermally broadened spectrum (within a radius $R_{\tau/2}=33.5$ kpc),
 while the right axis in {\it red} shows the flux of the direct, but {\it unobserved} spectrum (estimated from Hopkins et al. 2007).
 The colours of the axis correspond to those of the respective spectra. The line on the left is \ly, while the one
 to the right is Ly$\beta$ plus other lines in the Ly series.}
  {\includegraphics[width=0.5\textwidth \hspace{-3mm}]{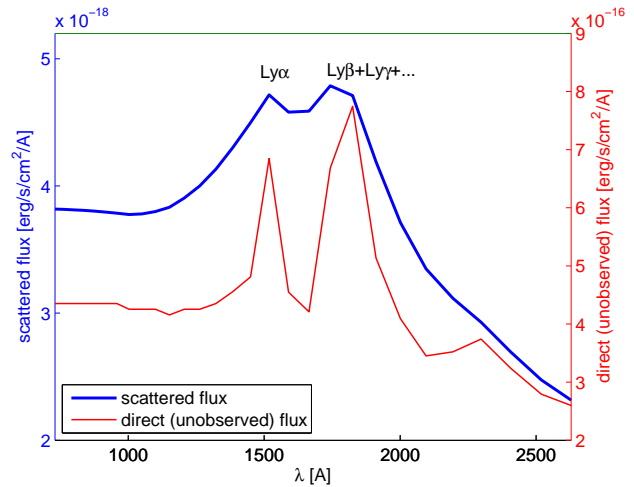}}
\label{fig:AGN_spec}
\end{figure}

Note that the important difference between the two cases discussed is that in the previous section (Perseus cluster) the source of emission is a diffuse region
(though small when compared to cluster scale, $R_c$) while in the present section (3C 295) the emission is from a point source, i.e. the central AGN.
For Perseus cluster there is an additional, but small, \ha flux arising from the AGN located in the central galaxy NGC 1275,
 on top of the dominant contribution from the central galaxy and the filaments around it.

\begin{figure}
  \centering
  \subfloat[Intensity of the scattered \ha ({\it solid thin red line}) and \ly ({\it solid thick blue line}) as a function of
 the projected distance $R$ from the cluster centre. The {\it dashed} lines in corresponding colours show the surface
 brightness profiles of the central galaxy 3C 295 estimated at the redshifted wavelength corresponding to the \ha/Ly$\alpha$ line.]
 {\includegraphics[width=0.5\textwidth \hspace{0mm}]{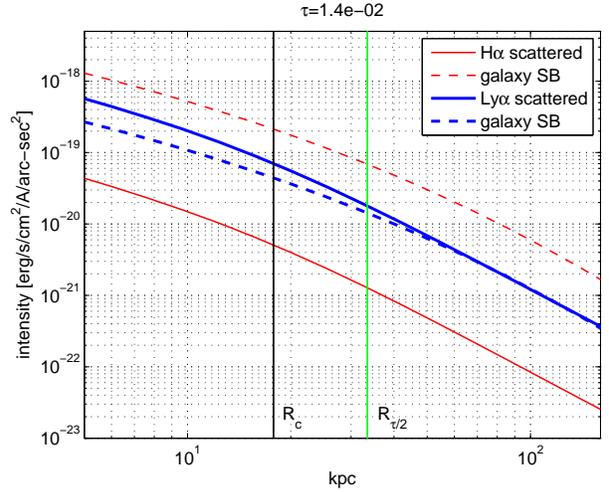}}
 \hspace{.03mm}
 \subfloat[Surface brightness (ABmag/arc-sec$^2$) of the scattered line of \ha ({\it solid thin red line}) and \ly ({\it solid thick blue line}) as a function of
 the projected distance $R$ from the cluster centre at the surface. The {\it dashed} lines in corresponding colours shows the surface
 brightness profiles of the central galaxy 3C 295 estimated at the redshifted wavelength corresponding to the \ha/Ly$\alpha$ line.]
 {\includegraphics[width=0.5\textwidth]{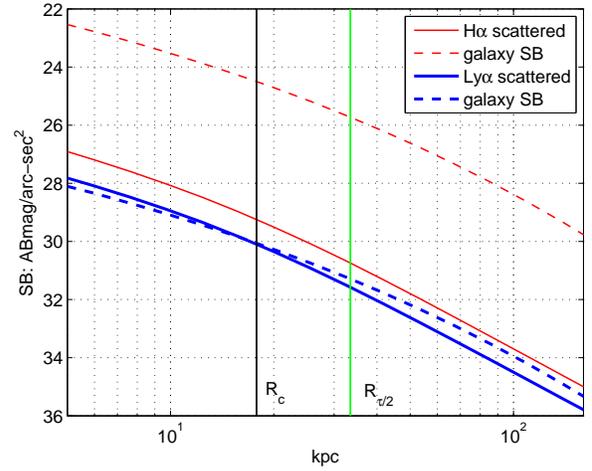}}\\
\subfloat[Expected polarisation of the scattered signal as a function of the projected radius for \ly ({\it thick blue}) and \ha ({\it thin red})
 scattered light.]
     {\includegraphics[width=0.5\textwidth]{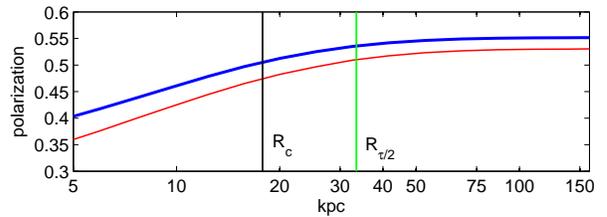}}
  \caption{Estimate of the expected signal of the diffuse scattered
    light in the galaxy cluster containing the central radio galaxy 3C 295 with a bright AGN. For the above plots $R_c$ and $R_{\tau/2}$ denote
 the core radius and the distance where the optical depth for the cluster falls to half its value. For this cluster, $1''$  corresponds to 5.8 kpc.}
  \label{fig:3c295}
\end{figure}

We now present estimates of the surface brightness of the scattered lines \ly and \ha of the AGN
 located within a rich cluster of galaxies. The observed hard band (2.0 - 10 keV) X-ray flux of the AGN in 3C 295, corrected for absorption,
 is $ 2.1 \times 10^{-13} {\rm erg/cm^2/s}$ \citep{2001MNRAS.324..842A}. This can be converted to an estimate for the bolometric
 luminosity using equation 19 of \cite{2012ApJ...757..181S}. \cite{2007ApJ...654..731H} provide a code that is used to obtain a template
 spectra as shown in Fig. \ref{fig:AGN_spec}. In Fig. \ref{fig:3c295} (a) and (b) we show the scattered intensity/surface brightness near the
lines \ha and \ly \footnote{\alert{In the spectrum of 3C 295 we should also directly observe narrow Ly$\alpha$, Ly$\beta$ etc. lines
 with fluxes $\sim$ 1-10\% (as is typical of AGN) of the corresponding (unobserved) broad lines.}}. 
For this we used the observed surface brightness profiles in the g and r bands \citep{2008MNRAS.389.1637B} and used the
average colour-magnitude relations for early-type galaxies from \cite{2005ApJ...622..244W} and \cite{2009MNRAS.398.2028J} to estimate the
 corresponding profiles in the bands relevant to the redshifted \ly and \ha lines. The galaxy surface brightness profiles as shown in the
 figures are estimated under the assumption that the de Vaucouleurs profile (which is seen to provide a good fit in the r and g bands)
 holds true across all the filters. However this is only an approximation since the radial colour profiles show a variation of about
 $\sim$ 1-2 magnitudes from galaxy to galaxy \cite[see for e.g.][]{2009MNRAS.398.2028J}, especially in the outskirts.

Thus we see that the scattered signal from central AGN in the \ly and \ha lines is expected to much stronger
 when compared to the previous case discussed in section \ref{sec:perseus}. In fact the surface brightness of this signal
is now comparable to the galaxy brightness, especially so for the \ly line.

\section{Feasibility of measuring the faint and diffuse scattered spectra}
\label{sec:feas}
The Thomson scattering emission, at the level of $\tau_T \sim 0.01$ of
the direct line flux is smeared out over the region $\sim
R_{\tau/2} \sim 20-50 ~ {\rm kpc}$ and forms a very smooth and faint diffuse
component. The surface brightness of this component falls with radius
as $\displaystyle I(R)\propto \frac{n_e(R)R}{R^2}\propto n_e(R)/R
\propto R^{-\alpha}$, where $\alpha=1.5-2$ over the radial range of
interest. Direct imaging of this faint component is extremely
challenging. However, one can use the spectral information on broadened
lines to separate this component from other components. One possible
strategy is to obtain a spectrum averaged over an area (extending typically up to tens of kpc in size),
 avoiding regions, contaminated by bright
filaments and/or the brightest parts of the optical galaxy. Here the
fact that the scattered signal is polarised  up to $\sim$ 50 \% (see Fig. \ref{fig:perseus} (c))
  could also be used to separate the scattered component from the direct component
 which is expected to be unpolarised.

In recent years there has been significant progress in observing faint diffuse objects. For example, the imaging of intracluster light in Virgo cluster
 by \cite{2005ApJ...631L..41M} using the 0.6 m Burrell Schmidt Telescope, and follow-on observations where they achieved a surface brightness of
 $\mu_V  \approx$ 29.2 \citep{2010ApJ...720..569R}. Another example is the imaging of the extended stream of stellar debris left over from the tidal
 disruption of a dwarf satellite galaxy by \cite{2008ApJ...689..184M} achieved using the 0.5 m Ritchey-Chrétien telescope of the BlackBird Remote Observatory.

In general, the challenges associated with such faint observations are many-fold \cite[see also][]{2011EAS....48..345D}:
\begin{enumerate}
 \item The sky brightness even at the best locations of ground-based telescopes is $\mu_V \approx 21.8$ (for moonless conditions). This is already 2--3 orders of
 magnitude brighter than the limits reached in above mentioned observations. The sky background must be carefully subtracted, including the effects
 of background variations, in order to prevent any confusion of the expected signal with the residuals. Of course this will not be an issue for imaging using
 space telescopes.

\item In any CCD detector there are always variations either in the pixel-to-pixel sensitivity or due to distortions in the optical path (e.g. vignetting
 produced at the periphery). In a perfect flat-fielded image, a uniform source is expected to produce a uniform image. The imaging of extremely faint objects
 depends crucially on the ability to achieve super-flat fields to be able to distinguish between real features in the image and residual artifacts. A
 flat-fielding is performed by combining various sky-patches which are slightly offset from the main image.

\item For the imaging of extremely faint objects the internal reflections from lenses and other parts of the telescope cannot be neglected. Their contribution
 can be reduced firstly by baffling of the telescope, keeping the number of optical elements (which act as reflecting surfaces) to the minimum and by using
 anti-reflective coatings on the optical surfaces to reduce the intensity of the reflections. Secondly, by modelling the complex pattern of internal
 reflections in software, and then removing them during post-processing analysis \citep{2009PASP..121.1267S}. In this context the use of a closed tube
 telescope is preferred over the open tube design.

\item Some other issues are the presence of halos, and their reflections from internal instrument optics, around bright foreground stars.
 These halos can be significantly brighter than the low surface-brightness background, which implies that subtraction will not be trivial.
Also the presence of galactic cirrus clouds along the line of sight will block and scatter the light from background objects.

\end{enumerate}

Very recently there have been immensely promising results reaching extremely low surface brightness up to 32 mag/arc-sec$^2$
 \cite[see][]{2014arXiv1401.5473A, 2014arXiv1401.5467V}. This  was made possible by the use of multiple commercially available
 Canon 400 mm f/2.8 IS II telephoto lenses having special anti-reflective coatings that can reduce the effect of scattered light and internal
 reflections by as much as a factor of 10. By using these small aperture lenses flat-fielding errors can be reduced to less than 0.1\%.
Secondly, by doing away with secondary mirror assembly the complex and variable pattern of diffraction and ghosting is suppressed.
These authors argue that by using a larger grid of multiple lenses, thus giving a larger {\it effective} aperture, it would be possible
 to make even further improvements.

Measuring this faint scattered light would certainly benefit from using spatial, spectral and polarisation information (see the beginning of sec. \ref{sec:feas})
 of the signal. The scattered component is expected to have a smooth profile both in wavelength and spatial extent, see Fig. \ref{fig:perseus}.
For this reason, neither a high spectral resolution nor highly accurate imaging should be essential (moderate pixel calibration/flat-fielding).
In the regions of interest measuring the average scattered light (for example, using stacking) over a small field of
 view ($\lesssim 1$') should be adequate. To measure the broadened line widths narrow band ($\sim$ 10--100 A) filters or a
 degraded spectrum from a multi-object spectrograph (MOS), used at wavelengths around the intrinsic bright line, should also suffice.
For narrow band filters, absolute flux calibration between the filters used at various wavelengths would be important to measure the shape
 of the broadened lines accurately. At the same time it would be important to avoid the bright regions and filaments.

Using a very high spatial resolution ($\sim$ 10 mas) would further help to exclude pixels containing stars and their halos. Obviously, this
 (ability to resolve individual stars) seems difficult and could be possible only for nearby galaxies/groups. Here, the sensitivity of the detector must
 be such that there is no confusion between the stellar light and the thermal noise in the CCD pixels; also the read out noise of the detector
 must be close to zero.

\section{Discussion}
\label{sec:discuss}

Bright emission lines are observed from the central galaxies of
cooling flow clusters. The surrounding ICM at temperatures
of a few keV with Thomson optical depth of $\sim 10^{-2}$ should scatter
the lines causing their broadening of order $\Delta \lambda / \lambda
\sim 15\%$. Therefore diffuse scattered light around the central galaxy can
be a potential diagnostic of the properties of ICM with regard to
electron temperature and density. The ratio of the continuum flux from the direct
light to the total\footnote{Alternatively, one could compute the optical
 depth as a function of radius by assuming, say a $\beta$-model, electron density
 profile and by measuring the scattered continuum flux within a given aperture at
 various projected radii.}
 scattered light should give us an estimate of the optical depth of the
cluster. The broadening of the bright emission lines can be used to
measure the temperatures. The fluxes are very faint and the detection
 of scattered emission is challenging.

We can also use the scattered lines in \ly and \ha as a probe of the ICM properties
in which the central AGN sits. The advantage of this method is that the luminosity
 of the line is at least 1-2 orders of magnitude higher when compared to the emission
 from the central galaxies in cooling core clusters. \ly line is better suited than \ha,
 as it is not only brighter but also the background galaxy surface brightness is much
 fainter in UV than in IR. The disadvantage is that one would require space based telescopes
to observe this signal. In addition the UV detector technology lags behind that in optical.
To give a perspective about the sensitivity required of an UV telescope to measure scattered \ly line
in the AGN of the central galaxy 3C 295, it would take $\sim$ 10 million seconds of observation with
 the ACS/SBC instrument\footnote{This exposure time estimate was derived assuming the F165LP filter
 and using the online exposure time calculator available at \url{http://etc.stsci.edu/etc/input/stis/imaging/}
for a S/N ratio of 3.0 in the imaging mode with 1$\times$1 pixel for the object diameter of 1$''$.}
 on the HST. Obviously, new technology is required to be able to measure this faint signal in a
 reasonable amount of time.

As the scattered lines would be highly broadened the spectra is expected to be
 extremely smooth. Hence, even observations from multiple narrow-band filters
covering the wavelengths of interest should suffice. See for example
 the ongoing survey J-PAS \citep[][]{2012MNRAS.423.3251A}
which would have 42 filters each having a bandwidth of $\sim 100 ~ {\rm A}$ starting from the UV band at 3500 A
up to the infra-red band at 10000 A.

The optimal cluster properties related to measurement of the scattered
light would be a large optical depth and the presence of bright (with
large equivalent width) emission lines in the central galaxy. It is
also preferable to have lower temperatures of the ICM so that the
features of the line are not completely washed out due to excessive
broadening. Some examples of clusters with these properties are
Zw3146 (z=0.29), A1835 (z=0.25), PKS 745-191 (z=0.10), RX J1347.5-1145
(z=0.45), RX J1532.9+3021 (z=0.36) and SPT-CL J2344-4243 (z=0.60).
To be able to mask individual
 pixels having stellar contribution (as outlined in section \ref{sec:feas})
 closer objects are better, since the surface brightness fluctuations
 decrease with distance. The scattered light should also be visible around large elliptical
galaxies, with much sharper features in the scattered lines due to
lower temperatures $T_e \sim 1$ keV, although with somewhat smaller
(by a factor of $\sim$3) optical depth.

The light scattered by the gas in the ICM is polarised and one
may in principle use this information to separate the contribution of
the direct stellar component which would be unpolarised. However,
polarisation measurements of such faint and diffuse light are expected
to be highly challenging even with the next-generation
instrumentation.

The measurements of optical depths along with X-ray measurements can
be used to obtain distances to clusters \citep{1982SvAL....8..175S,
  1990ApJ...363..344W}.  This is because both these measurements are
proportional to the LOS, but weighted differently ($\tau \sim \int n_e
dl$ and $S_X \sim \int n^2_e dl$). Alternatively the width of the
broadened lines is $\propto \int \sqrt{T} dl / \int dl$, and can be used
to probe the LOS temperature fluctuations, along with a knowledge of
the cluster size (using measurements of $y$ and optical depths for
example). For optical lines discussed above, the peak of the scattered lines will be Doppler shifted
depending on the LOS velocity of the gas at which the scattering
occurs and also depending on the relative velocity between the emitting
 regions and the scattering electrons. This offers a way to probe the
 transverse motions of gas and to understand the velocity structure
 in cluster cores \citep{2006MNRAS.367..433H, 2007MNRAS.380...33H, 2012ApJ...746..153M}.
 Additionally, the proposed method may provide a unique possibility to detect
 hot intrucluster gas around (relatively) distant type II quasars, since such hot gas may
 be difficult to find in X-rays because of the strong (in X-rays) central AGN and small angular size of the cluster.
And finally, the scattered light reaching us would be delayed with respect to the direct emission
by $\sim 10^5$ years. Therefore scattered emission reflects
mean luminosity of the line over the same time scale.

\section{Conclusions}
\label{sec:conclude}

The main conclusions of this paper are as follows:

\begin{enumerate}
\item The spectra of the central galaxies of clusters often show a few
   very bright emission lines. The detection of the Thomson
   scattered line emission by the intracluster gas will open new possibilities to
   probe the properties of the ICM using UV/optical/IR bands. The width
   of the broadened lines depends on the electron temperature while the
   ratio of the scattered light to the direct light probes the electron density (optical depth).

\item Due to the large thermal broadening the scattered spectrum would be very smooth.
Thus even a very low resolution spectrum or multiple narrow-band filters should suffice.
The scattered light needs to be collected from clusters typically within an area corresponding
 to a few 10's of kpc in radius from the centre. Depending on the redshift of the object this
 translates to distances of a few arc-mins to a few arc-secs. Although objects at higher redshifts
 suffer from low photon count rate and loss of photon energy, the latter might be compensated by
 the gain in the detector sensitivity at the redshifted wavelengths. For objects at higher redshifts
 there would not be any disadvantage from having a smaller field of view to measure the integrated
 scattered light.

\item This scattered light from the central galaxies in cool core clusters is expected to be very faint.
 Measurement of this faint signal needs an accurate understanding of the systematic effects relating to internal reflections in telescopes, flat-fielding and subtraction of
 sky brightness (for ground based telescopes). This appears to be extremely challenging even with the next generation instruments.

\item A more promising prospect seems to be to observe the scattered bright lines like \ly and \ha emitted by the central AGN sitting within
 rich galaxy clusters. The advantage of this method is that now the emission lines from the central AGN are much brighter.

\item Very recent imaging observations reaching 32 mag/arc-sec$^2$ using a grid of commercially available telephoto lenses appear to be highly
 promising with regards to the ability to reach extremely low-surface brightness, which would be useful for our purpose.

\item Since the scattered signal is expected to be very faint, it would be very useful to use the information from the spatial shape, profile
 and polarisation properties of the scattered component to separate it out from the much brighter direct stellar component.

\item Observations of the scattered spectrum can be used to measure distances to clusters, to measure the line-of-sight temperature fluctuations, and to probe the
variability of emission from the BCG on time scales of $\sim 10^5$ years.

\end{enumerate}

\section{Acknowledgements}

SK would like to thank Sandesh Kulkarni for many insightful conversations during the preliminary stages of this work.
 \alert{This research was partially supported by the Russian Foundation for Basic Research (grant 13-02-12250-ofi-m).}

\bibliographystyle{mn2e}
\bibliography{line}

\bsp
\label{lastpage}

\end{document}